\pdfoutput=1

\documentclass[11pt]{article}

\usepackage[]{acl}

\usepackage{enumitem,kantlipsum}
\usepackage{times} 
\usepackage{latexsym}

\usepackage[T1]{fontenc}

\usepackage[utf8]{inputenc}

\usepackage{microtype}

\usepackage{inconsolata}

\usepackage{graphicx}
\usepackage{stfloats}
\usepackage{multirow}

\usepackage{xcolor}

\usepackage{tabularx}
\usepackage{makecell}
\usepackage{booktabs}
\usepackage{ragged2e}
\usepackage{listings}
\usepackage[normalem]{ulem}
\useunder{\uline}{\ul}{}

\makeatletter
\def\blfootnote{\xdef\@thefnmark{}\@footnotetext}
\makeatother

\title{The Computational Anatomy of Humility: \\Modeling Intellectual Humility in Online Public Discourse}

\author{
        Xiaobo Guo$^{1*}$ \quad Neil Potnis$^{2*}$ \quad Melody Yu$^2$ \\
         \textbf{Nabeel Gillani}$^{2\spadesuit}$ \quad \textbf{Soroush Vosoughi}$^{1\spadesuit}$\\     
        $^{1}$Department of Computer Science, Dartmouth College, Hanover, New Hampshire\\
        $^{1}$\{xiaobo.guo.gr, soroush.vosoughi\}@dartmouth.edu\\
        $^2$Department of Art and Design, Northeastern University, Boston, Massachusetts\\
        $^2$\{n.potnis, yu.melo, n.gillani\}@northeastern.edu
    }

\begin{document}
\maketitle
\begin{abstract}
The ability for individuals to constructively engage with one another across lines of difference is a critical feature of a healthy pluralistic society. This is also true in online discussion spaces like social media platforms. To date, much social media research has focused on preventing ills---like political polarization and the spread of misinformation. While this is important, enhancing the quality of online public discourse requires not just reducing ills but also promoting foundational human virtues. In this study, we focus on one particular virtue: ``intellectual humility'' (IH), or acknowledging the potential limitations in one's own beliefs. Specifically, we explore the development of computational methods for measuring IH at scale. We manually curate and validate an IH codebook on 350 posts about religion drawn from subreddits and use them to develop LLM-based models for automating this measurement. Our best model achieves a Macro-F1 score of 0.64 across labels (and 0.70 when predicting IH/IA/Neutral at the coarse level), higher than an expected naive baseline of 0.51 (0.32 for IH/IA/Neutral) but lower than a human annotator-informed upper bound of 0.85 (0.83 for IH/IA/Neutral). Our results both highlight the challenging nature of detecting IH online---opening the door to new directions in NLP research---and also lay a foundation for computational social science researchers interested in analyzing and fostering more IH in online public discourse.\footnote{Our dataset is available at \href{https://github.com/xiaobo-guo/The-Computational-Anatomy-of-Humility-Modeling-Intellectual-Humility-in-Online-Public-Discourse}{Dataset}}
\end{abstract}

\blfootnote{
\hspace{-16.5pt}
$*$ Co-first author. \\
$\spadesuit$ Co-corresponding author.}

\section{Introduction}
The promise of social media, in many ways, was that it would untap our virtuous selves: our desire and ability to seek new knowledge and form connections across cultural and social divides. In some cases, this has happened, yet discourse on social media has also created an environment that often rewards the forgetting of our virtuous selves---leading to the spread of misinformation~\cite{vosoughi2018false}, embedding users within echo chambers~\cite{gillani2018socialmirror}, and leading individuals to harbor negative affect towards those who hold different beliefs~\cite{iyengar2019affective}. Many efforts to foster more constructive discourse online focus on addressing these ills, for example, by developing interventions that try to reduce affective polarization~\cite{saveski2022ff}, curtail the spread of misinformation~\cite{pennycook2021shifting}, and increase ``healthy'' politically cross-cutting exposures~\cite{gillani2018socialmirror,saveski2022engaging,santoro2022conversation,levy2021affective}. Yet few focus on cultivating deeper-seated human virtues that might prevent such ills from manifesting in the first place.

Intellectual humility is one such virtue and involves acknowledging the potential limitations of one's own beliefs, in part by welcoming diverse perspectives \cite{porter2017humility, whitcomb2017humility}. Greater intellectual humility has been associated with higher scrutiny of misinformation \cite{koetke2021covid}, less political ``my side'' bias \cite{bowes2021humility}, and less effective polarization \cite{bowes2020affective}: the degree to which individuals dislike political outgroups \cite{iyengar2019affective}.

Contemporary literature on intellectual humility (IH) has mostly been published in the social sciences and philosophy, exploring it as a cognitive phenomenon and thereby developing taxonomies and other measurement frameworks through questionnaire-based behavioral tasks \cite{leary2018IHreport}. Yet fostering greater IH on social media requires first developing scalable methods for detecting its presence or absence.

This paper introduces a novel task within the field of natural language processing: the automated detection and classification of intellectual humility (IH) and its opposite, intellectual arrogance (IA), in online public discourse. By developing methods to automatically detect IH and IA, we aim to deepen our understanding of which discussions—and participants—typically exhibit these traits. This insight is crucial for designing and deploying interventions that promote greater intellectual humility online. Such interventions are similar to those targeting related issues like polarization and civil discourse \cite{gillani2018socialmirror,santoro2022conversation,argyle2023assistant}. This research offers a fundamental step towards enhancing the quality and constructiveness of online interactions.

We present a detailed development of a comprehensive codebook and ontology for this task, accompanied by a systematic annotation process to refine these tools. Our study evaluates off-the-shelf large language models (LLMs) and exposes the substantial challenges these models encounter in accurately identifying IH and IA. Related work includes \citet{abedin2023exploring}, which estimates IH for study participants using standard psychological assessments and then analyzes language patterns from participant journals to identify predictors of IH. However, our study diverges significantly in its objectives and methods. Specifically, we employ more recent advances in NLP, focus on a different prediction task—namely, the multi-class classification of existing written content based on a predefined taxonomy rather than predicting IH scores generated by participants—and our research is set in the context of online Reddit discussions.

Specifically, our findings indicate that while human annotators achieve strong agreement, with a Cohen's Kappa of 0.67, the performance of LLMs, as evaluated by a Macro-average F1 score of 0.64 using GPT-4-turbo-2024-04-09, falls significantly short of human levels. Even treating the detection of Intellectual Humility as a three-class classification problem (IH/IA/Neutral) achieves a relatively low Macro-F1 score of 0.7, which falls significantly below a human annotator score of 0.83. Various advanced prompting methods, including Chain of Thought and automatic prompting, did not markedly improve model performance. These results highlight the promise of using computational methods to assess the presence or absence of intellectual humility at scale but also illustrate how doing so can be challenging, opening the door to new directions for research at the intersection of natural language processing and computational social science. Our specific contributions include:

\begin{enumerate} [leftmargin=*]
\itemsep0em 
 \item \textbf{Development of an Intellectual Humility Codebook:} We synthesized insights from social science and philosophy to create a framework tailored for analyzing online religious discourse. This codebook serves as a foundational tool for operationalizing the assessment of intellectual humility.
 \item	\textbf{Iterative Annotation Process:} Two trained annotators applied this codebook to 350 discussion threads from various religious subreddits, refining our approach through iterative coding to ensure robustness and repeatability of our annotations.
 \item	\textbf{Benchmarking LLM-Based Classifiers:} We assessed the capability of existing LLMs, particularly using the GPT-4-turbo-2024-04-09 model, in automating the detection of IH and IA. Our benchmarks reveal the current limitations of LLMs and underscore the necessity for developing more sophisticated models.

\end{enumerate}

\begin{figure*}[!t]
    \centering
    \includegraphics[width=2.0\columnwidth]{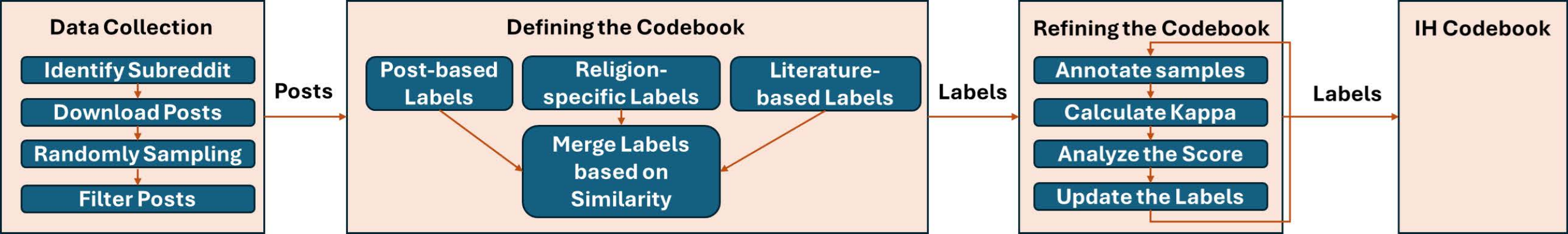}
    \caption{The flow chart for developing our IH Codebook}
    \label{fig:flow chart}
\end{figure*}
\section{Related work}

\paragraph{Religion and Humility}
Psychologists and philosophers have found that perceiving greater intellectual humility in one’s self---as well as one’s opponent in a religious conflict---is positively associated with the ability to forgive the offending party \cite{zhang2015religious}. This may, in part, underlie another finding in the religious domain: that greater intellectual humility can be positively associated with religious tolerance \cite{hook2017religious}. The latter study also identifies a positive correlation between intellectual humility and exposure to diversity, such that those reporting greater IH are more likely to express religious tolerance when exposed to diverse groups. This finding is underscored by recent work on related topics \cite{evolvi2018islamophobia}, and adds nuance to existing theories extolling the value of diverse exposures \cite{pettigrew2006contact}: such exposures may have a substantive positive impact on downstream outcomes of interest (like religious tolerance) when participants approach them with greater intellectual humility. This is particularly crucial online, where evidence supports that poorly presented exposure to diverse views can worsen divisions by amplifying extreme opinions from different sides rather than bridging gaps \cite{bail2018opposing, wsj2020polarization}. 

Intellectual humility in the religious domain is also relevant to interactions between religious and non-religious people. Contemporary discourse highlights that science and religion are often pitted against one another \cite{dennet2006religion}, despite the fact that religious or spiritual practices may help augment scientifically grounded practices like counseling \cite{gordon2018counseling} or offer ways of interpreting scientific phenomena \cite{lightman2023transcendant}. With diverse discourse on religion becoming more prevalent across online settings \cite{saha2023hate, evolvi2018islamophobia, parthasarathi2022covid, evolvi2019islam}, where the large scale of datasets often require computational modeling and analyses, there is a growing need to develop a computational infrastructure to both measure and enhance intellectual humility in digital discourse. 

\paragraph{Promoting Constructive Discourse Online}
Recent experimental studies have assessed how fostering constructive discourse online can be accomplished through tool-based interventions. Specifically, scholars have identified how ranking algorithms used by current social media platforms may contribute to greater affective polarization. Many experts suggest that implementing mechanisms to diversify the content consumed by users could reduce this polarization \cite{levy2021affective}. Researchers studying misinformation have found that prompting social media users to consider accuracy before sharing articles can reduce the spread of false information \cite{pennycook2021shifting}. Similarly, ``inoculating'' users against lower-quality content through light-touch coaching interventions has also been proven effective \cite{roozenbeek2022inoculation}. Some have even designed randomized interventions that connect people from opposing political groups to facilitate synchronous dialogue—finding that such dialogue can reduce affective polarization, but generally when discussing non-controversial topics \cite{santoro2022conversation}. 

More recent work \cite{argyle2023assistant} has explored how generative LLMs may assist participants involved in contentious conversations in rephrasing their contributions to be less toxic and more civil—pointing towards potential methods for fostering enhanced inclusivity in public settings. While these are important studies, most focus on reducing existing ills on social media platforms instead of seeking to promote virtues whose lack may be at the root of such ills to potentially produce more sustainable positive transformations in discourse and behavior~\cite{bowes2023intellectual}.

\section{Developing an IH Codebook}

Figure~\ref{fig:flow chart} illustrates the three steps we utilize in developing the IH Codebook, which include collecting data from Reddit, generating the initial Codebook based on LLMs and prior research, and annotating the samples as well as refining the Codebook.
    
\subsection{Data Collection}
We focus on discussions on Reddit centered around religious topics. This is chosen due to religion's enduring influence on global sociopolitical dynamics, notably in shaping policy and public dialogue in the United States and beyond. Furthermore, the inherently subjective nature of religious ``truth'' underscores the necessity for intellectual humility in fostering meaningful exchanges across diverse beliefs. Specifically, our study utilizes the r/ReligionHub's \emph{Directory of Religion-Related Subreddits}\footnote{\url{https://www.reddit.com/r/ReligionHub/comments/kohy3/directory_of_religionrelated_subreddits/}}, which lists 48 religion-oriented subreddits,  to define our dataset. From these, we downloaded all posts and comments spanning from June 2005 to December 2023 using the ``Academic Torrents'' archive.

In each subreddit, we randomly selected up to 500 posts and their associated comments, ensuring a broad representation of discussions. To avoid biases introduced by highly active users, we excluded posts from individuals who have made over 10,000 contributions to any single subreddit, as their views might not reflect those of the general user base. After preprocessing and applying these criteria, our dataset consisted of 17,160 posts. This collection was then used to develop and refine an IH taxonomy. 

\begin{figure*}[!htb]
    \centering
    \includegraphics[width=1.95\columnwidth]{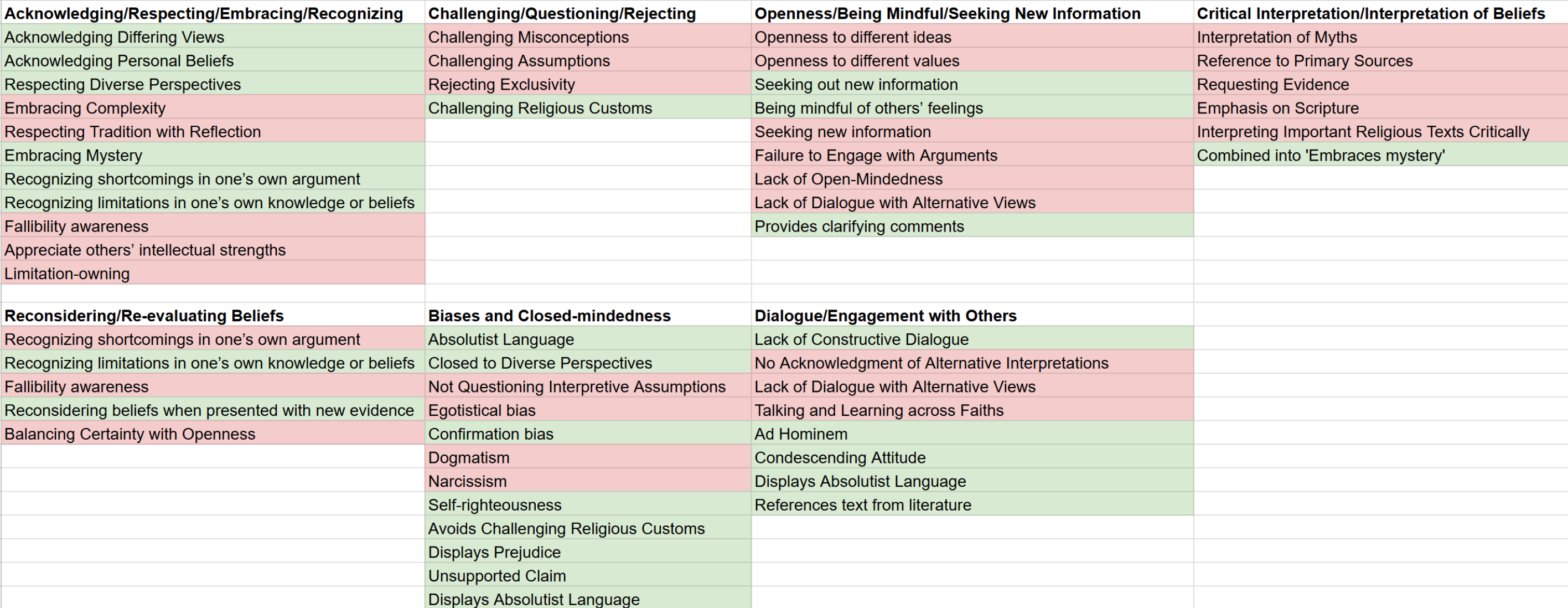}
    \caption{Manually identifying and eliminating similar labels for broader terms. The terms highlighted in green were then added to the first iteration of the codebook. The terms highlighted in red were the ones eliminated.}
    \label{fig: code group}
\end{figure*}

\subsection{Defining the Codebook}
We started assembling the codebook by first prompting ChatGPT (i.e., GPT 3.5) with a random sample of posts from religious subreddits and asking it to 1) classify each post as "intellectually humble" or "not intellectually humble", and 2), offer a justification for its classification. These outputs were then manually categorized into 21 broad categories: 12 reflecting intellectual humility (IH) and 9 reflecting its opposite---which, for convenience, we describe as ``intellectual arrogance'' (IA). The same exercise was repeated, but this time, GPT was prompted to provide classifications and justifications that were more specific to religion---asking it to indicate whether or not posts demonstrated ``religious intellectual humility.'' This exercise yielded 14 additional categories: 7 reflecting IH and another 7 reflecting IA. 
 
Next, we turned to existing literature describing different features of intellectual humility \cite{porter2022predictors, whitcomb2017humility, leary2017cognitive, porter2017humility}. Porter’s and Leary’s work on taxonomizing intellectual humility served as the main source for curating IH categories, which we condensed into shorter labels to include in the codebook \cite{porter2022predictors, whitcomb2017humility, leary2017cognitive}. This resulted in 20 additional terms: 12 reflecting IH and 8 reflecting IA.

We then used ChatGPT to identify similarities between the 55 terms and definitions and combine labels into a smaller codebook to begin applying to the Reddit posts through manual annotation. Combining labels and consolidating the codebook yielded a v1 codebook consisting of 25 labels: 12 reflecting IH and 13 reflecting IA. Figure~\ref{fig: code group} shows the full list of the original 55 labels. Red indicates labels that were removed or consolidated into one of the retained green labels.

\subsection{Refining the Codebook}
Two human annotators iteratively applied the codebook to annotate a random sample of Reddit posts (see Appendix~\ref{Appendix: human annotators} for the background of the annotators). We sampled up to 40 threads (if available) from each subreddit, yielding over 1,400 sampled threads across the 48 subreddits. For each thread, we also randomly determined whether the first or second comment (following the submission text that accompanied the start of the thread) would serve as the basis for annotation. In cases where the first comment was selected as the annotation target, only the submission text was chosen as context for that annotation; in cases where the second comment was selected, both the submission text and first comment served as context. 

Out of the 1,400 threads, human annotators read and collectively annotated 350 across a series of annotation waves (the remaining were not annotated due to the time intensiveness and complexity of the annotation task). Each annotation wave included the following steps: 1) selecting a subset of 50 to 100 posts to annotate (after reading the relevant post context), 2) after annotating, computing inter-annotator agreement (average Cohen's Kappa~\cite{mchugh2012interrater} across all applied codes), 3) analyzing per-code Kappa values to understand which codes had more or less agreement, 4) discussions between annotators to align on or refine codebook definitions before the next annotation wave. 

Cohen's Kappa values of 0.41-0.6 typically indicate moderate agreement; 0.61-0.8 indicate substantial agreement; and 0.81-1.0 indicate almost perfect agreement~\cite{mchugh2012interrater}. Annotators engaged in four successive annotation waves, with the first three involving 100 posts and the last one involving 50. The first wave was conducted with ongoing discussion and communication between annotators, yielding a Cohen's Kappa of 0.62. Annotations in the next wave were performed independently, yielding a much lower score of 0.35. After this, the annotators discussed disagreements and adjusted the codebook, obtaining average Kappa values of 0.6 and 0.7 in the remaining waves. It is important to note that the annotators did not revisit and re-annotate data from previous waves.

As the waves progressed, step 4 yielded several adjustments to the codebook. These adjustments typically involved eliminating or consolidating codes or updating code definitions. The decision to make these adjustments was typically made based on three factors: 1) significant overlap between the application of certain labels (like demonstrating self-righteousness and a condescending attitude, which often occurred together); 2) poorly-defined codes, particularly those reflecting a lack of some activity (like ``Avoids Challenging Religious Customs''---a post not challenging religious customs wouldn't necessarily mean it is demonstrating IA), and 3) related to 2), insufficient demonstration of intellectual humility or arrogance.

We computed a final average of Cohen's Kappa across all data from the various annotation waves using only the codes in the final codebook, yielding a score of 0.67. This score indicates strong agreement. We opted for this more conservative method rather than reporting only the highest Cohen's Kappa, which was obtained after our final wave. This approach ensures a more comprehensive representation of the kappa score, encompassing all data used to create the gold standard dataset. Many codes that led to lower agreement scores in earlier waves were either eliminated or merged as the annotation process progressed. Consequently, while the final kappa score is higher than those of earlier waves, it remains below the score achieved in the last wave, as expected.

\begin{table}[!htb]
    \centering
    \setlength\tabcolsep{3pt}
    \begin{tabular}{lccc}
    \hline
                        & Mean   & Std    & Max     \\ \hline
    \# Unique Labels    & 1.16   & 0.69   & 4.00    \\
    \# Context Words    & 266.44 & 344.66 & 2699.00 \\
    \# Context Sentence & 14.47  & 16.90  & 129.00  \\
    \# Target Words     & 98.03  & 16.90  & 129.00  \\
    \# Target Sentence  & 5.85   & 6.78   & 58.00   \\ \hline
    \end{tabular}
\caption{The descriptive statistics for the dataset include the mean, standard deviation (Std), and maximum (Max) number of unique labels per sample (\# Unique Labels), as well as the number of words in the context (\# Context Words), the number of sentences in the context (\# Context Sentences), the number of words in the target comment (\# Target Words), and the number of sentences in the target comment (\# Target Sentences).}
\label{tab: dataset descriptive}
\end{table}

Across all waves, 194 of the 350 posts were assigned at least one code from the final codebook. Table~\ref{tab:ih_codebook} summarizes the final codebook and inter-annotator agreement across codes. Some labels, such as ``Displays Prejudice'', appear infrequently in our datasets due to the limited scale. However, their importance to the concept of intellectual humility made it essential to retain them, as they address aspects specific to protected groups—like race, gender, and religion—that are not covered by other categories such as ``condescending attitude'' and ``ad hominem''. In addition to categorizing by individual labels, we assigned a composite ``IH/IA/Neutral'' label to all 194 posts. This binary classification was inferred from the cumulative presence of IH-related versus IA-related labels: a post with a predominance of IH labels was classified as IH, and vice versa for IA. Of the total, 134 posts were classified as IH, of which 5 included at least one IA label.
Conversely, of the 60 posts classified as IA, one included an IH label. There were no posts with an equal number of IH and IA labels. Table~\ref{tab: dataset descriptive} presents the descriptive statistics of our dataset.

Appendix~\ref{Appendix: samples for each label} displays sample comments corresponding to each label in the final codebook.

\linespread{1}
\begin{table*}[!htb]
\small
\begin{tabular}{llllcc}
\hline
Label                                                                                                       & IH/IA & Definition                                                                                                                                                                                  & Kappa & \# Agreed & \# Samples \\ \hline
\begin{tabular}[c]{@{}l@{}}Acknowledges Personal \\ Beliefs (APB)\end{tabular}                              & IH    & \begin{tabular}[c]{@{}l@{}}Affirms individual convictions with \\ the recognition that they are personal\\  perspectives, open to interpretation.\end{tabular}                              & 0.65  & 33        & 62         \\ \hline
\begin{tabular}[c]{@{}l@{}}Respects Diverse Perspectives \\ (RDP)\end{tabular}                              & IH    & \begin{tabular}[c]{@{}l@{}}Acknowledges a different perspective \\ in one's statement, and gives it \\ consideration and value.\end{tabular}                                                & 0.49  & 15        & 42         \\ \hline
Embraces Mystery (EM)                                                                                       & IH    & \begin{tabular}[c]{@{}l@{}}Accepts and appreciates the unknown \\ or spiritual aspects beyond full \\ comprehension.\end{tabular}                                                           & 0.66  & 4         & 8          \\ \hline
\begin{tabular}[c]{@{}l@{}}Recognizes limitations in \\ one’s own knowledge or \\ beliefs (RL)\end{tabular} & IH    & \begin{tabular}[c]{@{}l@{}}Understands that personal religious \\ knowledge or beliefs might not be \\ complete or fully accurate.\end{tabular}                                             & 0.70  & 10        & 18         \\ \hline
\begin{tabular}[c]{@{}l@{}}Reconsiders beliefs when \\ presented with new evidence \\ (RB)\end{tabular}     & IH    & \begin{tabular}[c]{@{}l@{}}Willingness to rethink religious \\ beliefs when faced with new \\ information that challenges them.\end{tabular}                                                & 0.80  & 4         & 6          \\ \hline
\begin{tabular}[c]{@{}l@{}}Seeks out new information \\ (SO)\end{tabular}                                   & IH    & \begin{tabular}[c]{@{}l@{}}Actively looks for new knowledge \\ and perspectives about different \\ religions or clarification on \\ statements made.\end{tabular}                           & 0.71  & 18        & 31         \\ \hline
\begin{tabular}[c]{@{}l@{}}Mindful of others’ feelings \\ (MF)\end{tabular}                                 & IH    & \begin{tabular}[c]{@{}l@{}}Considers how religious discussions or\\ actions might affect others emotionally.\end{tabular}                                                                   & 0.64  & 17        & 34         \\ \hline
\begin{tabular}[c]{@{}l@{}}Displays Absolutist Language \\ (DAL)\end{tabular}                               & IA    & \begin{tabular}[c]{@{}l@{}}Uses rigid language implying there's \\ only one absolute truth in religion.\end{tabular}                                                                        & 0.73  & 7         & 12         \\ \hline
\begin{tabular}[c]{@{}l@{}}Closed to Diverse \\ Perspectives (CDP)\end{tabular}                             & IA    & \begin{tabular}[c]{@{}l@{}}Unwillingness to consider, engage, \\ or accept viewpoints different from \\ one's own in religion.\end{tabular}                                                 & 0.66  & 7         & 14         \\ \hline
Condescending Attitude (CA)                                                                                 & IA    & \begin{tabular}[c]{@{}l@{}}Arrogant or dismissive behavior that\\  undermines others' perspectives or \\ intellect.\end{tabular}                                                            & 0.73  & 18        & 30         \\ \hline
Ad Hominem (AH)                                                                                             & IA    & \begin{tabular}[c]{@{}l@{}}The argument attacks the person \\ making the argument instead of \\ addressing the argument itself.\end{tabular}                                                & 0.87  & 7         & 9          \\ \hline
Displays Prejudice (DP)                                                                                     & IA    & \begin{tabular}[c]{@{}l@{}}Unfair opinions or judgments about \\ someone or a group without proper \\ understanding, often based on factors \\ like race, religion, or gender.\end{tabular} & 0.66  & 2         & 4          \\ \hline
Unsupported Claim (UC)                                                                                      & IA    & \begin{tabular}[c]{@{}l@{}}Assertion that lacks evidence or \\ adequate support, making it \\ unreliable or unverifiable.\end{tabular}                                                      & 0.45  & 3         & 10         \\ \hline
\end{tabular}
\caption{Final intellectual humility codebook (abbreviations per code are included in parentheses). ``Kappa'' indicates the computed Cohen's Kappa for each label across all annotation waves, ``\# Agreed'' indicates how many posts were assigned each label by both annotators, \# Samples shows the number of samples for each label in our dataset.}
\label{tab:ih_codebook}
\end{table*}
\vspace{-5pt}
\linespread{0.9}

\section{Evaluating LLMs for Automated IH Labeling}
      We utilized our developed codebook to conduct experiments with LLMs, aiming to assign IH and IA codes to posts using various prompt designs and model configurations. To ensure determinism in the generated responses, we set the temperature parameter of the LLMs to zero. This practice is recommended to ensure that the generated text is more focused and deterministic.

        \subsection{Experiment Settings}
       Our experiments involved querying the model to determine whether a post exhibits characteristics of IH, IA, or Neutral (``\textbf{IH/IA/Neutral''}) based on the taxonomy outlined in the codebook.

       We used the human-annotated data as a standard for evaluation, calculating the Macro-F1 score for model predictions. The F1 score reported is an average derived from comparing the labels generated by two annotators on a subset of samples and labels. It's important to note that annotators do not directly label the ``IH/IA/Neutral'' classification; instead, it is inferred from the aggregation of individual labels—if a post receives more IH-related than IA-related labels, it is classified as IH, and conversely as IA. Samples without either an IH or IA code are labeled as Neutral.

        Each experiment also includes a calculation of a baseline and an upper-bound F1 score: (1) The Baseline Macro-F1 score represents the expected performance of a naive model that randomly assigns labels based on the distribution of codes in the human-annotated data. (2) The inter-annotator agreement determines the Upper Bound. We treat the labels from Annotator A as the reference set to calculate the Macro-F1 score for Annotator B and vice versa. The upper bound is the mean of these two scores, providing a benchmark for the maximum achievable performance by the models under ideal inter-annotator consensus.

    \subsection{Prompt Design and Evaluation}\label{sct: prompt design}
        Prompt design is critical when conducting experiments with LLMs. To explore the impact of various prompt styles, we manually created multiple prompts, differing in both content and format.
       
        \textbf{Content Variations:} (1) \emph{Code-Only (C):} The prompt includes only the name of the code. (2) \emph{Description-Only (D):} The prompt includes only the description of the code, omitting its name. \emph{Code-and-Description (C\&D):} This comprehensive format includes both the name and the description of the code, allowing us to assess whether the integration of these elements influences model performance.

        \textbf{Format Variations:} (1) \emph{Multiple-Selection (MS):} This format presents all codes in a list, instructing the model to select codes that apply to the post. (2) \emph{Binary Question (BQ):} In this format, each code is considered individually; the model determines whether a specific code applies to a given post. We introduced these format variations to explore the primacy effects observed in LLMs, where models tend to prefer choices appearing earlier in a list~\cite{wang2023primacy,guo2024serial}. Note that for the IH/IA/Neutral prediction, it is impossible to format the question into \textit{BQ}; therefore, we only test with \textit{MS}.

        These variations resulted in six distinct prompt configurations. After the initial development, we iteratively refined the prompts to ensure the generated text adhered closely to the instructions. Detailed designs of these prompts can be found in Appendix~\ref{Appendix: prompts with different settings}.

        We assessed the effectiveness of these prompts using GPT-3.5-turbo-0125 and GPT-4-turbo-2024-04-09, the most recent versions of the GPT-3.5 and GPT-4 models at the time of our experiments. Table~\ref{tab: Task-wise IH prompt design performance} displays these prompts' performance metrics, detailing the average outcomes for all IH/IA/Neutral labels (``All'') and their efficacy in the IH/IA/Neutral binary classification.
        
\begin{table}[!htb]
\centering
\begin{tabular}{llll}
\hline
Prompt      & Model        & IH/IA/NE                       & All           \\ \hline
C-BQ        & GPT-3.5      & \multirow{2}{*}{0.62}          & 0.48          \\
C-MS        & GPT-3.5      &                                & 0.57          \\
C-BQ        & GPT-4        & \multirow{2}{*}{0.66}          & {\ul 0.63}    \\
C-MS        & GPT-4        &                                & 0.61          \\
D-BQ        & GPT-3.5      & \multirow{2}{*}{0.55}          & 0.50          \\
D-MS        & GPT-3.5      &                                & 0.58          \\
D-BQ        & GPT-4        & \multirow{2}{*}{{\ul 0.67}}    & 0.62          \\
D-MS        & GPT-4        &                                & 0.62          \\
C\&D-BQ     & GPT-3.5      & \multirow{2}{*}{0.59}          & 0.52          \\
C\&D-MS     & GPT-3.5      &                                & 0.59          \\
C\&D-BQ     & GPT-4        & \multirow{2}{*}{\textbf{0.70}} & \textbf{0.64} \\
C\&D-MS     & GPT-4        &                                & {\ul 0.63}    \\ \hline
Baseline    & Distribution & 0.32                           & 0.51          \\
Upper bound & Mutual       & 0.83                           & 0.85          \\ \hline
\end{tabular}
\caption{Performance metrics for IH and IA labels across various prompt designs. The table presents mean scores for all labels, labeled as ``All'' and the classification of samples as IH, IA or Neutral is indicated under the column ``IH/IA/NE''. The best performance for each label is highlighted in \textbf{bold}, while the second best is \underline{underlined}.}
\label{tab: Task-wise IH prompt design performance}
\vspace{-5pt}
\end{table}

        For the aggregated ``IH/IA/Neutral'' classification, the performance across all prompt designs is relatively uniform for each model and significantly exceeds the naive baseline. This uniformity suggests that the prompt design and model choice minimally impact the ``IH/IA/Neutral'' task outcomes. Conversely, for the Mean of Labels (``All''), we observe notable variability in results depending on the prompt design and model used. Generally, GPT-4 outperforms GPT-3.5 using the same prompt designs, with the Code-and-Description and Binary Question format (C\&D-BQ) yielding the best results across both tasks. For \textbf{Content Variations}, the Code-and-Description (C\&D) configuration consistently delivers superior performance for both models, likely due to its richer contextual input.
Regarding \textbf{Format Variations}, the Binary Question (BQ) format is more effective with GPT-4, while the Multiple-Selection (MS) format shows better results with GPT-3.5. This difference may stem from GPT-3.5's susceptibility to the ``primacy effect'', where altering the order of labels significantly impacts performance. However, it is unclear if this performance dip is solely due to primacy effects or if it is also influenced by changes in the prompt structure itself.

    Despite the LLMs' superiority over the naive baseline, a significant discrepancy remains between the models' performance and the human annotation-informed upper bounds, particularly in the task of label-wise prediction. The best-performing combinations of prompt and model fall short of the upper bounds by 0.13 for the ``IH/IA/Neutral'' task and 0.21 for ``Labels''. GPT-4 consistently outperforms GPT-3.5 across all labels, which is anticipated given GPT-4's larger model size. The Code-and-Description (C\&D) format achieves the highest scores for IH-specific labels, likely due to the richer context provided by these prompts. Detailed performance metrics for each label, excluding the overall mean performance across all labels, are presented in Appendix~\ref{appendix: label-wise performacne}.

        For IA-specific codes, the Description-only (D) strategy proves most effective. Conversely, the Code-and-Description (C\&D) format exhibits the weakest performance, suggesting that LLMs may process IA codes differently from IH codes. When synthesizing the results for both IH and IA codes, it is essential to recognize that performance can vary significantly among individual codes. The best-performing combinations of prompt and model still fall short of the upper bounds by 0.23 for the ``IH Mean'' and 0.15 for the ``IA Mean''.

        All these results, especially the label-wise results shown in Appendix~\ref{appendix: label-wise performacne}, underscore the inherent challenges in this task. Except for the methods based on LLMs, we also conducted experiments with classical methods on the ``IH/IA/Neutral'' task, utilizing TF-IDF or Bag-of-words for feature extraction followed by prediction based on the Logistic Regression. These methods perform similarly to our naive baseline based on the class distribution (with Macro-averaged F1 scores across five cross-validation folds of 0.36 and 0.39, respectively), highlighting the superiority of the LLM-based method. 

    \subsection{Interpretable Model Analysis}
        Following previous work~\cite{abedin2023exploring}, we show the top 5 positive/negative important words for IH, IA, and Neutral class. For this, we first utilize Logistic Regression with the TF-IDF feature to predict the coarse class (IH/IA/Neutral) and then utilize the Python ELI5 library for interpretable machine learning to understand the extent to which different features might influence the model's classifications. The keywords for each class are shown in Table \ref{tab: important words}.

        \begin{table*}[!htb]
        \centering
        \small
        \begin{tabular}{ll|ll|ll}
        \hline
        NE (Positive) & NE (Negative) & IA (Positive) & IA (Negative) & IH (Positive) & IH (Negative) \\ \hline
        https         & think              & based         & know          & case          & long          \\
        free          & god                & absolutely    & good          & religious     & did           \\
        important     & women              & think         & say           & mean          & https         \\
        going         & based              & children      & going         & ask           & doesn         \\
        years         & religion           & women         & time          & right         & books         \\ \hline
        \end{tabular}
        \caption{The top 5 positive/negative important words for IH, IA, and Neutral (NE), produced using Logistic Regression with TF-IDF features.}
        \label{tab: important words}
        \end{table*}

        We can observe that words that influence positive predictions in the model for the IH and IA classes demonstrate several patterns. For instance, the word ``absolutely'' tends to sway classifications towards IA, whereas less interpretable terms like ``https'' (perhaps indicating links to other resources) and ``did'' appear to reduce the likelihood of a piece of content being labeled IH. 

        In addition to this keyword-based analysis, in Appendix~\ref{Appendx: samples for different content varitions}, we demonstrate how label descriptions can impact model understanding, with LLMs sometimes mislabeling contextually tangible concepts like the ``Kingdom of Heaven'' as mystical. These discrepancies, possibly arising from ambiguities in defining IH or biases in the models, are further explored in Appendix~\ref{Appendix: chain of thoughts example}.

\subsection{Performance with Multiple Boost Methods}\label{sct: performance with boost methods}
    The experiments with various prompts and models underscore the challenges of label-wise prediction tasks. Given the significant performance gap between human annotators and our models, we implemented several boosting methods to enhance model performance, particularly using the C\&D-BQ settings with GPT-4, which provided the best initial results. We explored few-shot learning, chain-of-thought (CoT), automatic prompt optimization (Auto-Optimization), and iterative refinement with self-feedback (Self-Refinement). These methods are detailed in Appendix~\ref{app:boost} and ~\ref{Appendix: Boost Methods Prompts}. 
   
\begin{figure}[!hbt]
        \centering
        \includegraphics[width=\columnwidth]{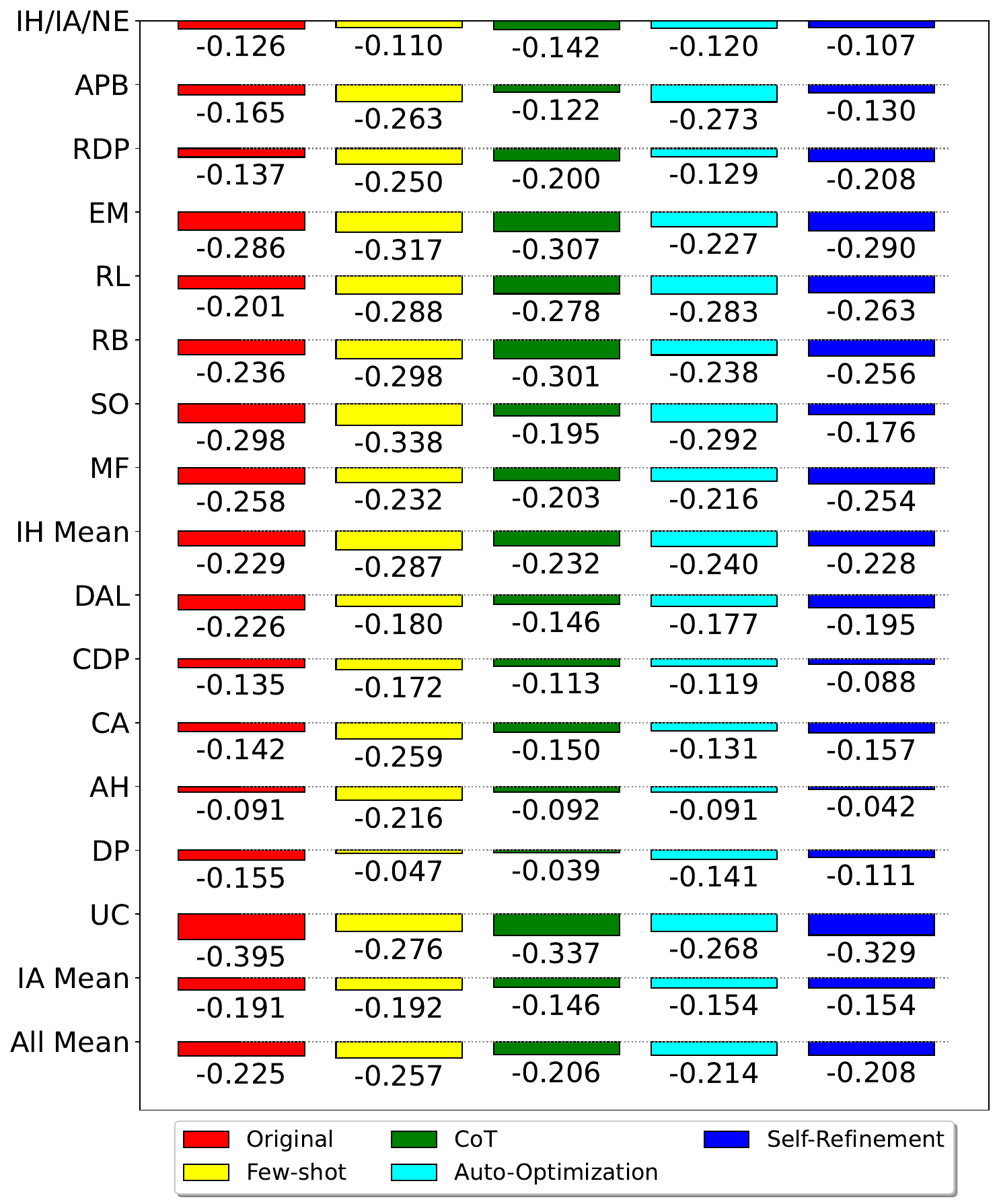}
        \caption{Comparison between different boosting methods and the human annotator upper bound; negative values indicate performance below the upper bound. ``Original'' refers to the results without any boosting.}
        \label{fig:boost_performance}
    \end{figure}

    Figure~\ref{fig:boost_performance} illustrates the impact of these methods compared to the human annotator's upper bounds, as detailed in Appendix~\ref{Appendix: boost method results}. All methods, except Few-shot, significantly improved performance on label-wise prediction tasks, achieving near-human levels for the "IH/IA/Neutral" (``IH/IA/NE'') task. The dip in performance observed in Few-shot learning is potentially due to overfitting, as discussed in~\citet{zhao2021calibrate}. The effectiveness of these methods varies by label. For IA labels, all methods typically surpass the original settings, while their impact on IH labels is less uniform. This suggests different underlying mechanisms in how LLMs process IH and IA labels. In general, considering the trade-off between performance and cost, no boosting method is suggested for this task, so prompting engineering should be more promising.

    Except for the GPT family, we also tested with other models, with outcomes detailed in Appendix~\ref{Appendix: various label results}.

    \section{Generalizability of the Dataset}
    The generalizability of the dataset is crucial to its utility and faces two main challenges: 1) extending the application of Intellectual Humility/Intolerance Ambiguity (IH/IA/Neutral) beyond religious contexts and 2) expanding the dataset's scope within the religious domain. These challenges are complicated by the reliance on expert judgment to identify IH/IA/Neutral, adding layers of complexity to both tasks.

    To broaden IH to additional domains, it is necessary to supplement the existing codebook with new, domain-specific codes. While many existing codes are adaptable across various contexts where IH is relevant, the codebook has primarily been developed focusing on religious codes that resonate across broader domains. For example, the principle of “respecting diverse perspectives” is crucial for nurturing constructive dialogue across religious divides and serves as a foundation of intellectual humility in wider discussions. Similarly, the concept of “embracing mystery,” though initially tailored for religious discourse, can be interpreted more broadly to encompass an appreciation for the unknown in conversations—a fundamental aspect of intellectual humility that acknowledges the limits of our knowledge.

    Addressing the dataset's size involves straightforwardly increasing the number of expert-annotated samples. However, training new experts presents challenges, given the specialized expertise required for precise annotation. Alternatively, considering the high quality of our current labeled data, we propose using our dataset for validation and utilizing labels generated by LLMs as training samples. This method employs techniques such as few-shot learning or classifiers fine-tuned with these samples for automated labeling. While this approach may introduce some noise, the integrity of the dataset is maintained through the quality of the initial annotations. Appendix~\ref{Appendix: samples generated by LLMs} displays examples of samples identified or generated by LLMs using this method.

\section{Conclusion and Future Work}

This paper introduces a methodology for the automated detection and classification of intellectual humility and its counterpart, intellectual arrogance, in online religious discourse. Our comprehensive codebook development and systematic annotation process underpins this novel task, providing a robust framework for operationalizing the nuanced measurement of these intellectual virtues and vices. Despite achieving significant inter-annotator agreement and developing advanced prompting strategies, our experiments reveal that current LLMs, including the state-of-the-art GPT-4, fall short of replicating human-like discernment in this complex domain. 

The findings from this research demonstrate the potential of LLMs to assist in the proactive enhancement of online discourse and expose the limitations inherent in current technology. These insights encourage further academic inquiry and collaboration to refine these models. 

Detecting intellectual humility in online discourse can be practically helpful in several ways. For example, the classifiers we develop may power future technological tools like~\cite{katsaros2022intervene} and others previously deployed on social media platforms to detect opportunities for discussion participants to enhance the IH of their posts and encourage them to do so before posting. Social media platforms or independent developers might also design LLM-powered tools that enable participants to update posts to help others feel more understood, akin to how intellectually humble posts might function. These applications, some of which we describe in the paper, could be pursued further through field experiments on platforms like Reddit or with standing panels of social media users. Defining and detecting an amorphous concept like intellectual humility is a starting point for building these tools and understanding where they need to be deployed in digital discourse settings.

\section{Limitations}

This research confronts several limitations that underscore the challenges inherent in the computational modeling of complex social concepts like Intellectual Humility and Intellectual Arrogance.

Firstly, defining IH and IA through a codebook, a necessary step for operationalization introduces a degree of abstraction from the more nuanced, real-world expressions of these virtues and vices. Our codebook captures features indicative of IH and IA within online discourse, but these indicators may only be approximate correlates of the underlying intellectual traits. Despite achieving substantial inter-annotator agreement, the variability in this agreement across different codes reflects the subjective nature of these constructs and the difficulty in attaining a universally accepted operational definition.

Moreover, the scale of our annotated dataset is another significant constraint. With only 350 posts evaluated, of which fewer than 200 were coded with IH or IA labels, our findings are based on a relatively small data pool. This sample size limits the generalizability of our conclusions and reflects the intensive nature of the annotation process, which can be both time-consuming and complex. Quality concerns drove our decision against using crowdsourced platforms like Prolific.com to obtain annotations; however, scaling up the dataset will necessitate finding a balance between data quantity and annotation quality.

Lastly, the performance of the LLMs used in this study may have been constrained by the designs of our prompts. Despite careful engineering, the prompts might not have adequately captured the complexity needed to elicit accurate discernment of IH and IA from the models. This limitation was observed across various model configurations and might have also impacted the efficacy of the automatic prompting methods.

These limitations highlight the need for ongoing refinement of both the methodological approaches and the theoretical frameworks used in studies of this nature. As we advance our understanding and techniques, we must continually evaluate and adapt our strategies to better capture the intricate dynamics of intellectual virtues in online communication.

\section{Ethics statement}

Our research is driven by the goal of promoting more respectful and open-minded discourse online. However, we recognize that the tools we develop for the scalable detection and measurement of Intellectual Humility could potentially be misused. There is a risk that these tools might be employed for censoring speech, enforcing uniformity in discourse, or other controlling measures that could undermine the diversity of perspectives essential for a pluralistic society. Such misuse would directly contradict our objective of cultivating a broad spectrum of views on social media, a goal that fundamentally requires intellectual humility and a readiness to embrace diverse viewpoints.

Furthermore, we must consider the inherently normative aspect of defining IH. Our codebook and the annotations it guides are influenced by a specific cultural and philosophical perspective, which may not universally capture the essence of IH or its antithesis, Intellectual Arrogance. There is also a risk that employing LLMs for automating IH detection could perpetuate existing biases or create illusions of understanding, as highlighted by \citet{kidd2023distort} and \citet{messeri2024illusion}. Such risks necessitate ongoing vigilance to ensure these tools do not simplify or distort complex interpersonal traits.

As IH classifiers evolve and potentially surpass current baselines, it is essential to continuously reflect on and refine what IH entails. Tools like the ones we introduce in this paper should be used diagnostically to enhance understanding and foster intellectual virtues rather than to dictate or limit the expression of ideas.

Ultimately, this project does not claim to offer a definitive method for defining or measuring intellectual humility. Instead, it aims to initiate the development of frameworks that can enhance the detection and, eventually, promotion of intellectual humility in online discourse settings. By providing these tools, we hope to support those committed to practicing and promoting intellectual humility, thereby enriching online discourse and contributing to a more thoughtful and tolerant virtual community.

\section*{Acknowledgments}
We are grateful to the Templeton Foundation and Georgia State University for supporting this work.

\bibliography{custom}

\setcounter{table}{0}
\setcounter{figure}{0}
\renewcommand\thefigure{\Alph{section}\arabic{figure}}
\renewcommand\thetable{\Alph{section}\arabic{table}}
\clearpage
\appendix
\colorlet{punct}{red!60!black}
\definecolor{background}{HTML}{EEEEEE}
\definecolor{delim}{RGB}{20,105,176}
\colorlet{numb}{magenta!60!black}
\lstdefinelanguage{json}{
    basicstyle=\small\ttfamily,
    numbers=left,
    numberstyle=\small,
    stepnumber=1,
    numbersep=8pt,
    xleftmargin=16pt,
    showstringspaces=false,
    breaklines=true,
    frame=lines,
    backgroundcolor=\color{background},
    literate=
     *{0}{{{\color{numb}0}}}{1}
      {1}{{{\color{numb}1}}}{1}
      {2}{{{\color{numb}2}}}{1}
      {3}{{{\color{numb}3}}}{1}
      {4}{{{\color{numb}4}}}{1}
      {5}{{{\color{numb}5}}}{1}
      {6}{{{\color{numb}6}}}{1}
      {7}{{{\color{numb}7}}}{1}
      {8}{{{\color{numb}8}}}{1}
      {9}{{{\color{numb}9}}}{1}
      {:}{{{\color{punct}{:}}}}{1}
      {,}{{{\color{punct}{,}}}}{1}
      {\{}{{{\color{delim}{\{}}}}{1}
      {\}}{{{\color{delim}{\}}}}}{1}
      {[}{{{\color{delim}{[}}}}{1}
      {]}{{{\color{delim}{]}}}}{1},
}

\section{IH Codebook Development}
        
    \subsection{Human Annotators} \label{Appendix: human annotators}
        There are two human annotators for the process of refining the Codebook. Here we provide the details of these two annotators:
        \paragraph{Annotator \#1:} Full-time Research Assistant with degrees in computational design (MSc) and arts technology (BS). Native English speaker with previous work experience in information science.
        \paragraph{Annotator \#2:} Undergraduate researcher studying computer science and philosophy, with a specific focus on intellectual humility.  Native English speaker. 

    \subsection{Samples for Each Label} \label{Appendix: samples for each label}
        Below, we present an example from the dataset, which includes the post's title, content, target comment, labels from two annotators (Labels\_1 and Labels\_2), and the IH/IA/Neutral label:
        \begin{itemize}
            \item \textbf{Post Title:} I am an Ásatrúar, I believe in the Aesir ; Vanir. AMA.
            \item \textbf{Content:} I am a Pagan, an Ásatrúar. I follow Odin, Loki, Thor, Freya, Frigg, Tyr, and many other gods and goddesses. They are called the Aesir and Vanir. I have been of this faith for a little over two years now, converted by my husband. Before that, I didn't really label myself. I've always known that something is out there, and had moments where I have felt the Gods presence around me.  I believe my Gods; Goddesses were once real people, who did great things. Thus were immortalized in the ancient Norse ""myths"". I do not believe the Gods to be omnipotent or anything of the sort, but I do believe they have influence over their ""aspects"". For example, Freya is the Goddess of Fertility(among other things), so if you were to pray to her and build her an altar, she would have the power from Valhalla to influence your chances of being fertile or becoming pregnant. Could also go for if you were attempting to have a garden, she could influence the fertility and growth of the garden.   Please, ask me anything. :)
            \item \textbf{Target Comment:} For what purpose would someone invoke or pray to Loki? My knowledge of Ásatrúar is limited to some of what i have read on wiki and basic stories i have heard.
            \item \textbf{Lables\_1:} Recognizes limitations in one’s own knowledge or beliefs, Seeks out new information
            \item \textbf{Labels\_2:} Recognizes limitations in one’s own knowledge or beliefs, Seeks out new information
            \item \textbf{IH/IA/Neutral Label:} IH
        \end{itemize}
    
        In Table~\ref{tab: Samples for each label of the IH codebook}, we show the target comment for each label in our Codebook including the label, the IH/IA/Neutral the code belonging to, and one example of each label. 
    \begin{table*}
    \small
    \centering
    \begin{tabular}{lll}
\hline
    Label                                                                                               &IH/IA           & Example                                                                                                                                                                                                                                                                                                                \\ \hline 
    Acknowledges Personal Beliefs                                                                       & IH             & \begin{tabular}[c]{@{}l@{}}\#71 “I was under the impression that the Tanakh \\ was closer to the biblical Old Testament than the Torah, \\ and that the Torah was included in the Tanakh. \\ Am I horribly wrong?”\end{tabular}                                                                                        \\ \hline 
    Respects Diverse Perspectives                                                                       & IH             & \begin{tabular}[c]{@{}l@{}}\#34 “Thanks!  We are an interdenominational school \\ serving both Protestants and Catholics.  \\ This will make a wonderful resource for making sure \\ I cover Catholic perspectives.  Great find!”\end{tabular}                                                                         \\ \hline 
    Embraces Mystery                                                                                    & IH             & \begin{tabular}[c]{@{}l@{}}\#112 “Everything is spiritual in nature. \\ When we are in a calm and meditative like state \\ I believe our energy, that came from nature, \\ can reach out and intermingle with the life energy\\ that is Gaia. …”\end{tabular}                                                          \\ \hline 
    \begin{tabular}[c]{@{}l@{}}Recognizes limitations in one’s \\ own knowledge or beliefs\end{tabular} & IH             & \begin{tabular}[c]{@{}l@{}}\#2 “For what purpose would someone invoke or pray to Loki? \\ My knowledge of Ásatrúar is limited to some of \\ what i have read on wiki and basic stories i have heard.”\end{tabular}                                                                                                     \\ \hline 
    \begin{tabular}[c]{@{}l@{}}Reconsiders beliefs when presented \\ with new evidence\end{tabular}     & IH             & \begin{tabular}[c]{@{}l@{}}\#115 “Well, what I've found at least in study with ancient Greeks \\ is that often sexuality for them was encouraged to be explored, \\ especially because the genders were often segregated at \\ young ages for a time. But I will definitely look more into things…”\end{tabular}       \\ \hline 
    Seeks out new information                                                                           & IH             & \begin{tabular}[c]{@{}l@{}}\#8 “Are there differences between the book of mormon that \\ the lds church uses and the one that you use?”\end{tabular}                                                                                                                                                                   \\ \hline 
    Mindful of others’  feelings                                                                        & IH             & \begin{tabular}[c]{@{}l@{}}\#226 “You aren't the first to make painful life altering mistakes. \\ We are only human. \\ Do what you can to make things right if it's possible.”\end{tabular}                                                                                                                           \\ \hline 
    Displays Absolutist Language                                                                        & IA             & \begin{tabular}[c]{@{}l@{}}\#69 “...Every other religious scripture is full of errors \\ but not the quran.   The quran is inimitable. …”\end{tabular}                                                                                                                                                                 \\ \hline 
    Closed to Diverse Perspectives                                                                      & IA             & \begin{tabular}[c]{@{}l@{}}\#57 - “Short version: God is good! Any criticism of Him \\ is false.  The Bible is true: Jesus is the ONE \\ and ONLY way to heaven. God loves you. Bless you! ”\end{tabular}                                                                                                              \\ \hline 
    Condescending Attitude                                                                              & IA             & \begin{tabular}[c]{@{}l@{}}\#69 “There is no doubt that Islam is the truth, \\ its evidences aren't based on dreams or a gut feeling. \\ We have tangible proofs and if you reject \\ we'll then that's up to you”\end{tabular}                                                                                        \\ \hline 
    Ad Hominem                                                                                          & IA             & \begin{tabular}[c]{@{}l@{}}\#63 “I don't think he's the worst person to ever live \\ I just don't think he was presidential. \\ He only won because Hillary Clinton is a scary bitch.”\end{tabular}                                                                                                                    \\ \hline 
    Displays Prejudice                                                                                  & IA             & \begin{tabular}[c]{@{}l@{}}\#303 “You dumb female atheists cannot even defend atheism \\ in a debate. You are all so irrational. And you are the arrogant ones. \\ You are so ignorant that you think figuring out god doesn't exist \\ makes you superior to theists but you really need to be humbled.”\end{tabular} \\ \hline 
    Unsupported Claim                                                                                   & IA             & \begin{tabular}[c]{@{}l@{}}\#260 “The hard truth that homophobic Christian’s \\ don’t want to admit is that most of the \\ “clobber verses” are mistranslated (on purpose) and \\ the original text never mentioned homosexuality.”\end{tabular}                                                                       \\ \hline
    \end{tabular}
    \caption{Examples for each label of the IH codebook}
    \label{tab: Samples for each label of the IH codebook}
    \end{table*}
\clearpage
\section{Details of the Boost Methods} \label{app:boost}
    \begin{itemize}[leftmargin=*]
    \itemsep0em 
        \item \textbf{Few-shot learning}: This method, which generally improves LLM performance, involves providing three positive and three negative samples related to each label.
        \item \textbf{CoT}: Prior experiments suggest that requiring LLMs to articulate their reasoning enhances performance. We opted for a 6-shot setup as preliminary tests indicated it offers optimal results, constrained by the available number of samples labeled ``DP''.
        \item \textbf{Auto-Optimization}: Adopting techniques from \citet{ye2023prompt}, we applied automatic prompt optimization to GPT-4. For each label, six samples and the current prompt were given to the model for analysis and updating. Over ten rounds, three new prompts were generated each round; the one showing the best performance was selected for subsequent rounds. This iterative process continued without significant overfitting issues.
        \item \textbf{Self-Refinement}: Based on \citet{madaan2024self}, this method uses prediction, feedback, and reconsider cycles to iteratively improve outputs. Incorporating the chain of thought, two rounds of refinement were conducted per sample, as additional rounds did not yield further improvements.
    \end{itemize}
\section{Prompts}
    \subsection{Prompts with Different Content and Format} \label{Appendix: prompts with different settings}
        \subsubsection{IH/IA/Neutral System Prompt}
        \begin{lstlisting}[language=json]
        {{#system~}}
        You are a classifier for predicting whether the given text is intellectual humility, intellectual arrogant, or neutral. You must choose the answer from the following options: neutral, intellectual humility, and intellectual arrogance.
        Intellectual humility means recognizing that their beliefs might be wrong, including the following features: {{IH_code}}. If it follows any of this, it should be labeled as Intellectual humility.
        Intellectual arrogance is a state of mind where someone has an exaggerated view of their own intellect and knowledge and believes it is superior to others, such as the following features: {{IA_code}}. The list of features is not exhaustive.
        Neutral means not related to religious discourse or not enough information to classify as intellectual humility or intellectual arrogance.
        {{~/system}}
        \end{lstlisting}
        where the \{IA\_code\} and \{IH\_code\} is the fine-grained label for IH and IA. This is the prompt for the setting of Code-and-Description, for the description-only just remove the parts about the fine-grained code, and for the description-only, we have the following prompt:

        \begin{lstlisting}[language=json]
        {{#system~}}
        You are a classifier for predicting whether the given text is intellectual humility, intellectual arrogant, or neutral. You must choose the answer from the following options: neutral, intellectual humility, and intellectual arrogant.
        Intellectual humility means that it follows at least one of the following features: {{IH_Code}}. If it follows any of this, it should be labeled as Intellectual humility.
        Intellectual arrogant means that it follows at least one of the following features:  {{IA_Code}}. If it follows any of this, it should be labeled as Intellectually arrogant.
        Neutral means not related to religious discourse or lacking sufficient information for classification.
        {{~/system}}
        \end{lstlisting}
        
        \subsubsection{Label-wise Prediction System Prompt}
        For the question about each code for IH and IA, we have two versions one is the Multiple-Selection and the other is binary-question. For the Multiple-Selection, the prompt is:
        \begin{lstlisting}[language=json]
{{#system~}}
    Your task is to label the given text from Reddit about religion. The given text includes the Title, the content of the Submission, the content of the Comment, and the Target Text. Please label the Target Text with one or more labels from the following list:"{{Code_list}}".
    Each sample might be labelled with multiple labels, please separate each label with
{{~/system}}
                                                   
        \end{lstlisting}
        where the \{Code\_list\} can be the list of code names, code descriptions, or the combination of them based on the settings of code-only, description-only, and code-and-description.

        For the Bianry-Question setting, we have one system prompt for each code. The prompt is as follows:
        \begin{lstlisting}[language=json]
{{#system~}}
    Your task is to label the given text from Reddit about religion. The given text includes the Title, the content of the Submission, the content of the Comment, and the Target Text. If the Target Text can be described as `{{Code}}`, answer `Yes`. If it does not fit this description, answer `No`. 
{{~/system}} 
        \end{lstlisting}
        where the \{Code\} can be the name of the code, the description of the code, or the combination of them based on the settings of code-only, description-only, and code-and-description.
    \subsubsection{User Prompt}
        For all tasks and settings, the user prompt is the same including the information of the input samples as follows:
        \begin{lstlisting}[language=json]
{{#user~}}
Here is a discussion with the title: '{{Post_title}}', and the content is as follows: '{{Post_content}}'. The first comment is: '{{First_comment}}'. 

{{#if focus_on_second_comment}}
    The second comment is: '{{Second_comment}}'. 
{{/if}}

Based on the content do you think Comment: '{{Focal_comment}}' is '{{Label}}' or not.
{{~/user}}
        \end{lstlisting}
        where \{Post\_title\}, \{Post\_content\}, \{First\_comment\}, and \{Second\_comment\} are the content from the samples,  \{focus\_on\_second\_comment\} is a mark to show whether we analyze the first or second comment, \{Focal\_comment\} is the content of the analyzed comment, and the \{Label\} is the label to ask.
        
    \subsection{Prompts for Different Boost Methods}\label{Appendix: Boost Methods Prompts}
        \subsubsection{Prompts for Few-shot and Cot}
        In this section, we list all the prompts for the boost methods used in our paper. All the prompts are created based on the Coda-and-Description setting with Binary Question, which achieves the best performance for GPT-4.
        For the settings of Few-shot and CoT, here is the prompt we utilize:
        \begin{lstlisting}[language=json]
{{#system~}}
    {{System_Prompt}}
    {{#if CoT}}
        You must explain how you get the answer first then responding the answer.
    {{/if}}
{{~/system}

{{#if few_shot}}
    {{#user~}}
        {{User_Prompt}}
    {{~/user}}
    {{#assistent~}}
        {{Answer}}
    {{~/assistent}}
{{/if}}

{{#user~}}
    {{User_Prompt}}
{{~/user}}
            
        \end{lstlisting}
        where \{System\_Prompt\}, \{User\_Prompt\} is the one designed in the Appendix~\ref{Appendix: prompts with different settings}, \{CoT\} and \{few\_shot\} is the mark for labeling whether we utilize CoT and Few-shot respectively, and \{Answer\} is the results of the provided samples. We have 6 samples for the Few-shot, but we only show one sample here.
        
        \subsubsection{Prompts for the Auto-Optimization}
        For the Auto-Optimization, here is the prompt used for optimizing the prompts for the task:
        \begin{lstlisting}[language=json]
{{#system~}}
    {{System_Prompt}}
    You are a helpful assistant.
{{~/system}

{{#user~}}
    A prompt is a text paragraph that outlines the expected actions and instructs the model to generate a specific output. This prompt is concatenated with the input text, and the model then creates the required output.
                         
    In our collaboration, we'll work together to refine a prompt. The process consists of two main steps:
                         
    ## Step 1
    I will provide you with the current prompt, how the prompt is concatenated with the input text (i.e., "full template"), along with 6 example(s) that are associated with this prompt. Each examples contains the input, the reasoning process generated by the model when the prompt is attached, the final answer produced by the model, and the ground-truth label to the input. Your task is to analyze the examples, determining whether the existing prompt is decsribing the task reflected by these examples precisely, and suggest changes to the prompt.
                         
    ## Step 2
    Next, you will carefully review your reasoning in step 1, integrate the insights to craft a new, optimized prompt. Optionally, the history of refinements made to this prompt from past sessions will be included. Some extra instructions (e.g., the number of words you can edit) will be provided too.
{{~/user}}

{{#assistant}}
    Sure, I'd be happy to help you with this prompt engineering problem. Please provide me with the prompt engineering history, the current prompt, and the examples you have.
{{~/assistant}}

{{#user}}
    ## Prompt
    {{Curr_prompt}}
    
    ## Examples
    {{Examples}}

    ## Prompt Refinement History from the Past
    Note that higher accuracy means better. If some edits are useful in the past, it may be a good idea to make edits along the same direction.
    {{history_performance}}
    
    ## Instructions
    For some of these examples, the output does not match with the label. This may be due to the prompt being misleading or not describing the task precisely.
    
    Please examine the examples carefully. Note that the ground-truth labels are __absolutely correct__, but the prompts (task descriptions) may be incorrect and need modification. For each example, provide reasoning according to the following template:
    
    ### Example <id>
    Input: <input>
    Output: <output>
    Label: <label>
    Is the output correct compared to the label: <yes or no, and your reasoning>
    Is the output correctly following the given prompt: <yes or no, and your reasoning>
    Is the prompt correctly describing the task shown by the input-label pair: <yes or no, and your reasoning>
    To output the correct label, is it necessary to edit the prompt: <yes or no, and your reasoning>
    If yes, provide detailed analysis and actionable suggestions to edit the prompt: <analysis and suggestions>
    
    You must analyze all exaples provided.
{{~/user}}

{{#assistant}}
    {{Analysis}}
{{~/assistant}}

{{#user}}
    Now please carefully review your reasoning in Step 1 and help with Step 2: refining the prompt. 
    ## Current Prompt
    {Curr_Prompt}
    
    ## Prompt Refinement History from the Past
    Note that higher accuracy means better. If some edits are useful in the past, it may be a good idea to make edits along the same direction.
    {{Analysis}}
    
    ## Instructions
    * Please help edit the prompt so that the updated prompt will not fail on these examples anymore.
    * Reply with the prompt. Do not include other text.
{{~/user}}

{{#assistant}}
    {{Updated_prompt}}
{{~/assistant}}

{{#user}}
    Now please summarize what changes you've made to the prompt, in the following format. Make sure the summariy is concise and contains no more than 200 words.
    " * At step {}, the prompt has limitations such as <summary of limitations>. Changes to the prompt include <summary of changes>."
    Reply with the summarization. Do not include other text.
{{~/user}}

{{#assistant}}
    {{change_summary}}
{{~/assistant}}
          
        \end{lstlisting}
        where \{System\_Prompt\} is the one designed in the Appendix~\ref{Appendix: prompts with different settings}, \{Curr\_Prompt\} is the current optimized prompt which will be the user\_prompt designed in the Appendix~\ref{Appendix: prompts with different settings}, \{Examples\} are a list of 6 samples with 3 positive and 3 negative the same as CoT, \{history\_performance\} is a list of historical optimizing summary and the corresponding performance, \{Analysis\} is the analysis from the LLMs for the current prompt, and the \{change\_summary\} is a brief summary of the changes for this iteration generated by the LLM.
        
        \subsubsection{Prompts for the Self-Refinement}
        For the Self-Refinement, we adjusted the previous prompts. Here are the prompts we use to guide the models in generating feedback.
        
        For the task of ``IH/IA/Neutral'', we have:
        \begin{lstlisting}[language=json]
{{#system}}
You are an AI model providing feedback on a intellectual humility prediction task. The following code is used to describe not intellectual humility: 
{{IA_Code}}.
The following code is described as intellectual humility:
{{IH_Code}}
{{~/system}}

{{#user}}
{{User_Prompt}} Please provide a brief feedback especially challenging the existing result
{{~/user}}
        \end{lstlisting}
        where \{User\_Prompt\} is the one designed in the Appendix~\ref{Appendix: prompts with different settings}
        
        For the task of label-wise prediction, we have:
        \begin{lstlisting}[language=json]
{{#system}}
You are an AI model providing feedback on a prediction about whether the Given Text can be described as {{Code}}.
{{~/system}}

{{#user}}
{{User_Prompt}} Please provide a brief feedback especially challenging the existing result
{{~/user}}
        \end{lstlisting}
        where \{Code\} is the Code-and-Description of the label and \{User\_Prompt\} is the one designed in the Appendix~\ref{Appendix: prompts with different settings}
        
        Here are the prompts for reconsidering based on the feedback generated by the LLMs.

        For the task of ``IH/IA/Neutral'', we utilize the following prompts:
        \begin{lstlisting}[language=json]
{{#system}}
    You are an expert in the domain of intellectual humility. Your task is to label the given texts from Reddit about religion based on the provided feedback. Please label the given texts as intellectual humility or not intellectual humility with `No` or `Yes`. If you think the given text can be described as at least of the following
    {IA_Code}},
    you should answer `No`. If the given text cannot be described as at least of the following
    {{IH_code}}, 
    you should answer `Yes`.
    You must explain how you get the answer first then responding Yes or No.
{{~/system}}
{{#user}}
{{~/user}}
        \end{lstlisting}
        where the \{IA\_code\} and \{IH\_code\} is the Code-and-Description list of all labels.

        For the task of label-wise prediction, we have
        \begin{lstlisting}[language=json]
{{#system}}
    Your task is to label the given text from Reddit about religion. The given text includes the Title, the content of the Submission, the content of the Comment, and the Target Text. If the Target Text can be described as `{{Code}}`, answer `Yes`. If it does not fit this description, answer `No`. 
    You must explain how you get the answer first then responding the result.
{{~/system}}
{{#user}}
    Here is a discussion with the title: '{{Post_title}}', and the content is as follows: '{{Post_content}}'. The first comment is: '{{First_comment}}'. 
    
    {{#if focus_on_second_comment}}
        The second comment is: '{{Second_comment}}'. 
    {{/if}}
    
    Based on the feedback {{Feedback}} and the content do you think Comment: '{{Focal_comment}}' is '{{Label}}' or not.
{{~/user}}
        \end{lstlisting}
        where \{Code\} is the Code-and-Description of the label, \{Post\_title\}, \{Post\_content\}, \{First\_comment\}, and \{Second\_comment\} are the content from the samples,  \{focus\_on\_second\_comment\} is a mark to show whether we analyze the first or second comment, \{Focal\_comment\} is the content of the analyzed comment, the \{Label\} is the label to ask, and the \{Feedback\} is the generated feedback from the LLM.
        
    \begin{figure*}[!htb]
        \centering
        \includegraphics[width=1.95\columnwidth]{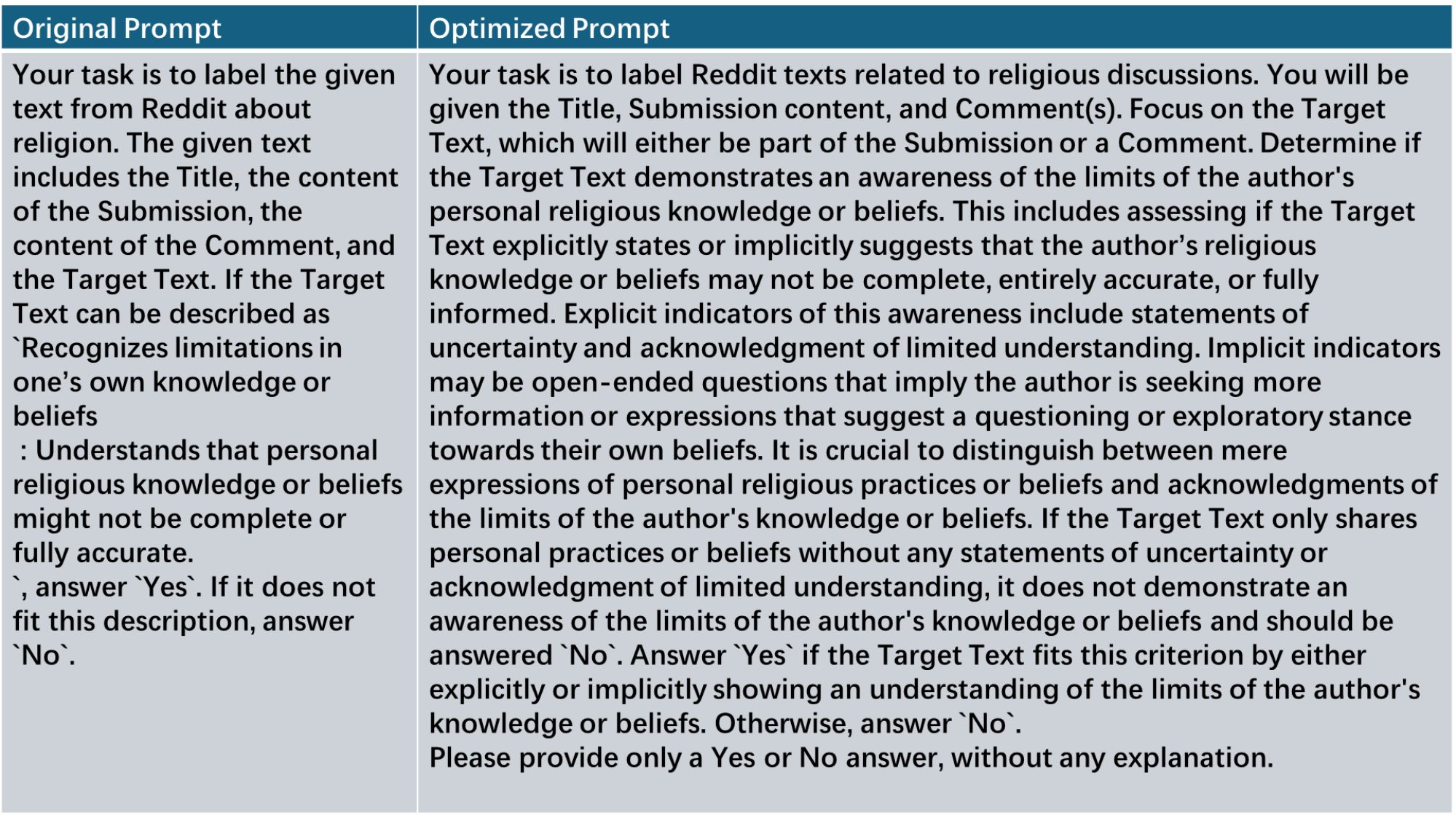}
        \caption{The original and optimized system prompts for BQ settings of the code ``Recognizes limitations in
one’s knowledge or beliefs''}
        \label{fig: optimized prompt}
    \end{figure*}        
        
    \subsection{Optimized Prompt}\label{Appendix: optimized prompt}
    In figure~\ref{fig: optimized prompt}, we show the original and the optimized prompts for the code ``Recognizes limitations in one’s own knowledge or beliefs''. As we can observe, compared with the original prompt, it includes much more data for guiding the model to emphasize on certain situations. This will increase the performance on certain samples while it will also decrease the performance when the samples are very different since it might guide the model wrongly. 

\section{Results}
    \subsection{Label-wise Performance of Various Content and Format} \label{appendix: label-wise performacne}
    Except for the aggregate performance of ``IH/IA/Neutral'' classification and the mean performance of all labels, Table~\ref{tab: Label-wise IH prompt design performance} detailed the performance of each label for all the combinations.
    
    \begin{table*}[!htb]
\centering
\begin{tabular}{ll|llllllll}
\hline
Prompt        & Model        & \multicolumn{8}{c}{IH Labels}                                                                                                 \\
              &              & APB           & RDP           & EM            & RL            & RB            & SO            & MF            & IH Mean       \\ \hline
C-BQ          & GPT-3.5      & 0.47          & 0.55          & 0.53          & 0.42          & 0.49          & 0.49          & 0.44          & 0.48          \\
C-BQ          & GPT-4        & 0.67          & 0.52          & 0.59          & {\ul 0.65}    & 0.56          & 0.59          & \textbf{0.74} & 0.62          \\
C-MS          & GPT-3.5      & 0.59          & 0.4           & 0.63          & 0.51          & 0.48          & 0.60          & 0.57          & 0.54          \\
C-MS          & GPT-4        & 0.63          & 0.55          & 0.63          & 0.55          & 0.5           & 0.60          & 0.62          & 0.58          \\ \hline
D-BQ          & GPT-3.5      & 0.60          & 0.23          & 0.50          & 0.53          & 0.47          & 0.56          & 0.47          & 0.48          \\
D-BQ          & GPT-4        & 0.69          & {\ul 0.62}    & 0.55          & 0.58          & 0.55          & 0.61          & 0.66          & {\ul 0.61}    \\
D-MS          & GPT-3.5      & 0.64          & 0.23          & 0.53          & 0.49          & 0.46          & {\ul 0.70}    & {\ul 0.69}    & 0.53          \\
D-MS          & GPT-4        & 0.57          & 0.5           & 0.49          & 0.58          & 0.57          & 0.67          & 0.64          & 0.57          \\ \hline
C\&D-BQ       & GPT-3.5      & 0.58          & 0.41          & 0.49          & 0.48          & 0.54          & 0.57          & 0.6           & 0.53          \\
C\&D-BQ       & GPT-4        & {\ul 0.70}    & \textbf{0.63} & 0.57          & \textbf{0.66} & {\ul 0.59}    & 0.62          & 0.63          & \textbf{0.63} \\
C\&D-MS       & GPT-3.5      & 0.65          & 0.41          & \textbf{0.71} & 0.47          & \textbf{0.61} & 0.58          & 0.60          & 0.58          \\
C\&D-MS       & GPT-4        & \textbf{0.73} & 0.57          & {\ul 0.64}    & 0.58          & 0.52          & \textbf{0.71} & 0.66          & \textbf{0.63} \\ \hline
Baseline      & Distribution & 0.54          & 0.48          & 0.48          & 0.58          & 0.48          & 0.5           & 0.52          & 0.51          \\
Upper bound   & Mutual       & 0.87          & 0.77          & 0.86          & 0.86          & 0.83          & 0.92          & 0.89          & 0.86          \\ \hline
Prompt Desgin & Model        & \multicolumn{8}{c}{IA Labels}                                                                                                 \\
              &              & DAL           & CDP           & CA            & AH            & DP            & UC            &               & IA Mean       \\ \hline
C-BQ          & GPT-3.5      & 0.46          & 0.38          & 0.52          & 0.68          & 0.41          & 0.39          &               & 0.47          \\
C-BQ          & GPT-4        & 0.69          & 0.63          & \textbf{0.75} & 0.73          & 0.62          & 0.52          &               & {\ul 0.66}    \\
C-MS          & GPT-3.5      & 0.58          & {\ul 0.69}    & 0.57          & 0.75          & 0.59          & 0.50          &               & 0.61          \\
C-MS          & GPT-4        & 0.66          & 0.65          & 0.68          & 0.73          & 0.61          & 0.56          &               & 0.65          \\ \hline
D-BQ          & GPT-3.5      & 0.54          & 0.44          & 0.63          & 0.61          & 0.41          & 0.46          &               & 0.51          \\
D-BQ          & GPT-4        & 0.67          & 0.66          & \textbf{0.75} & \textbf{0.80} & 0.61          & 0.37          &               & 0.64          \\
D-MS          & GPT-3.5      & \textbf{0.77} & 0.48          & 0.53          & 0.49          & \textbf{0.86} & \textbf{0.65} &               & 0.63          \\
D-MS          & GPT-4        & {\ul 0.75}    & \textbf{0.70} & 0.68          & 0.64          & {\ul 0.74}    & {\ul 0.60}    &               & \textbf{0.68} \\ \hline
C\&D-BQ       & GPT-3.5      & 0.53          & 0.45          & 0.55          & 0.69          & 0.45          & 0.46          &               & 0.52          \\
C\&D-BQ       & GPT-4        & 0.64          & 0.65          & \textbf{0.75} & {\ul 0.77}    & 0.59          & 0.44          &               & 0.64          \\
C\&D-MS       & GPT-3.5      & 0.65          & 0.67          & 0.58          & 0.54          & 0.63          & 0.53          &               & 0.60          \\
C\&D-MS       & GPT-4        & 0.62          & 0.65          & {\ul 0.72}    & 0.66          & 0.61          & 0.52          &               & 0.63          \\ \hline
Baseline      & Distribution & 0.47          & 0.51          & 0.52          & 0.48          & 0.49          & 0.49          &               & 0.49          \\
Upper bound   & Mutual       & 0.87          & 0.78          & 0.89          & 0.86          & 0.75          & 0.83          &               & 0.83          \\ \hline
\end{tabular}
\caption{Performance metrics for IH and IA labels across various prompt designs. The table presents mean scores for all Intellectual Humility (IH) and Intellectual Arrogance (IA) labels, labeled as ``IH Mean'' and ``IA Mean'' respectively. The best performance for each label is highlighted in \textbf{bold}, while the second best is \underline{underlined}}
\label{tab: Label-wise IH prompt design performance}
\end{table*}
    
    \subsection{Various Model Results}\label{Appendix: various label results}
        In extending our examination to other LLMs, such as Gemini-1.0-pro and Claude-3-opus-20240229, we applied the Code-and-Description Binary Question (C\&D-BQ) format, both with and without CoT prompting. These models performed optimally in setups tuned for GPT-4, as shown in Table~\ref{tab: prompt various model}

        \begin{table}[!htb]
\small
\begin{tabular}{l|ll|ll}
\hline
Code & \multicolumn{2}{l|}{Gemini-1.0} & \multicolumn{2}{l}{Claude-3} \\
     & CoT           & No Cot         & CoT          & No Cot        \\
    \hline
    IH/IA/Neutral   & 0.43          & 0.43        & 0.38      & 0.35              \\
    \hline
    APB  & 0.65          & 0.62           & 0.56         & 0.58          \\
    RDP  & 0.45          & 0.36           & 0.43         & 0.47          \\
    EM   & 0.51          & 0.57           & 0.52         & 0.51          \\
    RL   & 0.39          & 0.43           & 0.52         & 0.64          \\
    RB   & 0.53          & 0.52           & 0.49         & 0.50           \\
    SO   & 0.65          & 0.61           & 0.61         & 0.64          \\
    MF   & 0.62          & 0.61           & 0.57         & 0.57          \\
    IH Mean &0.53 &0.56 &0.54 &0.53                                      \\
    \hline
    DAL  & 0.64          & 0.62           & 0.68         & 0.63          \\
    CDP  & 0.62          & 0.62           & 0.6          & 0.65          \\
    CA   & 0.67          & 0.71           & 0.63         & 0.65          \\
    AH   & 0.65          & 0.69           & 0.68         & 0.67          \\
    DP   & 0.55          & 0.61           & 0.58         & 0.7           \\
    UC   & 0.43          & 0.51           & 0.4          & 0.44          \\
    IA Mean & 0.59       & 0.62           & 0.59      & 0.63          \\
\hline
\end{tabular}
\caption{The performance of Gemini-1.0 and Claude-3 with the settings C\&D-BQ with/without chain of thoughts.}
\label{tab: prompt various model}
\end{table}
        
        Analysis of Table~\ref{tab: prompt various model} reveals that Claude-3 and Gemini-1.0-pro underperform compared to GPT-4, likely due to the prompts being specifically optimized for GPT-4. Moreover, while CoT generally boosts the ``IH/IA/Neutral'' task performance for these models, it tends to diminish efficacy across other coding tasks. This behavior is also different from what we observed in the GPT-4 model. We are not sure what the cause of this observation is, but it also might be because of the prompt design.

    \subsection{Boost Method Results}\label{Appendix: boost method results}
        In Table~\ref{tab: boost performance}, we show the exact value of different boost method for GPT-4-turbo-2024-04-09.
                \begin{table*}[!hbt]
            \centering
            \begin{tabular}{lccccc}
            \hline
            Label        & Original & Few-shot & CoT   & Auto-Optimization & Self-Refinement \\ \hline
            IH/IA/Neutral& 0.704    & 0.720    & 0.688 & 0.710             & 0.723           \\ \hline
            APB          & 0.705    & 0.607    & 0.748 & 0.597             & 0.740           \\
            RDP          & 0.633    & 0.520    & 0.570 & 0.641             & 0.562           \\
            EM           & 0.574    & 0.543    & 0.553 & 0.633             & 0.570           \\
            RL           & 0.659    & 0.572    & 0.582 & 0.577             & 0.597           \\
            RB           & 0.594    & 0.532    & 0.529 & 0.592             & 0.574           \\
            SO           & 0.622    & 0.582    & 0.725 & 0.628             & 0.744           \\
            MF           & 0.632    & 0.658    & 0.687 & 0.674             & 0.636           \\
            IH Mean      & 0.631    & 0.573    & 0.628 & 0.620             & 0.632           \\ \hline
            DAL          & 0.644    & 0.690    & 0.724 & 0.693             & 0.675           \\
            CDP          & 0.645    & 0.608    & 0.667 & 0.661             & 0.692           \\
            CA           & 0.748    & 0.631    & 0.740 & 0.759             & 0.733           \\
            AH           & 0.769    & 0.644    & 0.768 & 0.769             & 0.818           \\
            DP           & 0.595    & 0.703    & 0.711 & 0.609             & 0.639           \\
            UC           & 0.435    & 0.554    & 0.493 & 0.562             & 0.501           \\
            IA Mean      & 0.639    & 0.638    & 0.684 & 0.676             & 0.676           \\ \hline
            All Mean     & 0.635    & 0.603    & 0.654 & 0.646             & 0.652           \\ \hline
            \end{tabular}
            \caption{The performance of different boost methods as well as the one without any boost method.}
            \label{tab: boost performance}
        \end{table*}
        
\section{Samples with Different Settings}
    \subsection{Samples for different Content Variations}\label{Appendx: samples for different content varitions}
        Here we show an example for showing the difference between the \textit{Code-Only}, \textit{Description-Only}, and \textit{Code-and-Description} with the Chain-of-Thought settings mentioned in Section~\ref{sct: performance with boost methods}, to show the difference caused by the inclusion of description:
        \begin{itemize}
            \item \textbf{Label: }Mindful of others’ feelings
            \item \textbf{Description: }Considers how religious discussions or actions might affect others emotionally.
            \item \textbf{Code-Only Reason: } “The target text is a question posed by someone who is curious about the practices of Ásatrúar, specifically regarding the purpose of invoking or praying to Loki. The commenter openly admits their limited knowledge, which is based only on what they have read on Wikipedia and basic stories they have heard. This admission of limited knowledge and the phrasing of the question in a non-confrontational manner indicates a genuine curiosity and a respectful approach to learning more about someone else's beliefs. The commenter's approach is mindful of others' feelings in several ways: 1. **Acknowledgment of Limited Knowledge**: By stating their limited understanding, the commenter shows respect for the complexity and depth of the Ásatrúar faith, implying that they do not wish to make assumptions or judgments based on incomplete information. 2. **Openness to Learning**: The question is framed in a way that invites explanation and dialogue rather than challenging or dismissing the beliefs being discussed. 3. **Respectful Curiosity**: The specific inquiry about Loki, a figure who might be controversial or misunderstood due to his portrayal in popular media and mythology, is asked in a manner that seeks to understand the religious and cultural significance rather than perpetuate stereotypes. Given these considerations, the target text can be described as "Mindful of others’ feelings" because it demonstrates a respectful and open-minded approach to discussing religious beliefs that are different from one's own. Therefore, the answer is **Yes**.”
            \item \textbf{Description-Only: } "The Target Text is a question from a commenter asking about the purpose of invoking or praying to Loki, a figure in the Ásatrúar faith. The commenter also mentions their limited knowledge of Ásatrúar, indicating they are seeking information or clarification. This text does not explicitly or implicitly consider how religious discussions or actions might affect others emotionally. Instead, it focuses on gaining understanding or information about a specific practice within a religion. There is no mention or implication of emotional impacts on others due to religious discussions or actions. Therefore, the answer is `No`."
            \item \textbf{Code-and-Description Reason: }“The Target Text is a comment asking a question about the Ásatrúar religion, specifically about the purpose of invoking or praying to Loki. The commenter acknowledges their limited knowledge about the religion, indicating a willingness to learn and understand more. However, there is no explicit or implicit consideration of how religious discussions or actions might affect others emotionally. The commenter is simply seeking information. No”
        \end{itemize}
        
        By comparing these generated texts, we observe that the code-only approach wrongly labels the text as it interprets “Mindful of others’ feelings” as “a respectful and open-minded approach to discussing.” The main issue is that the model generates a definition for the code based on its learned knowledge, which may not align with a general or expert understanding of the code. This highlights the importance of providing clear descriptions of the codes.

    \subsection{Chain of Thoughts Example}\label{Appendix: chain of thoughts example}
\begin{figure*}[!hbt]
            \centering
            \includegraphics[width=1.95\columnwidth]{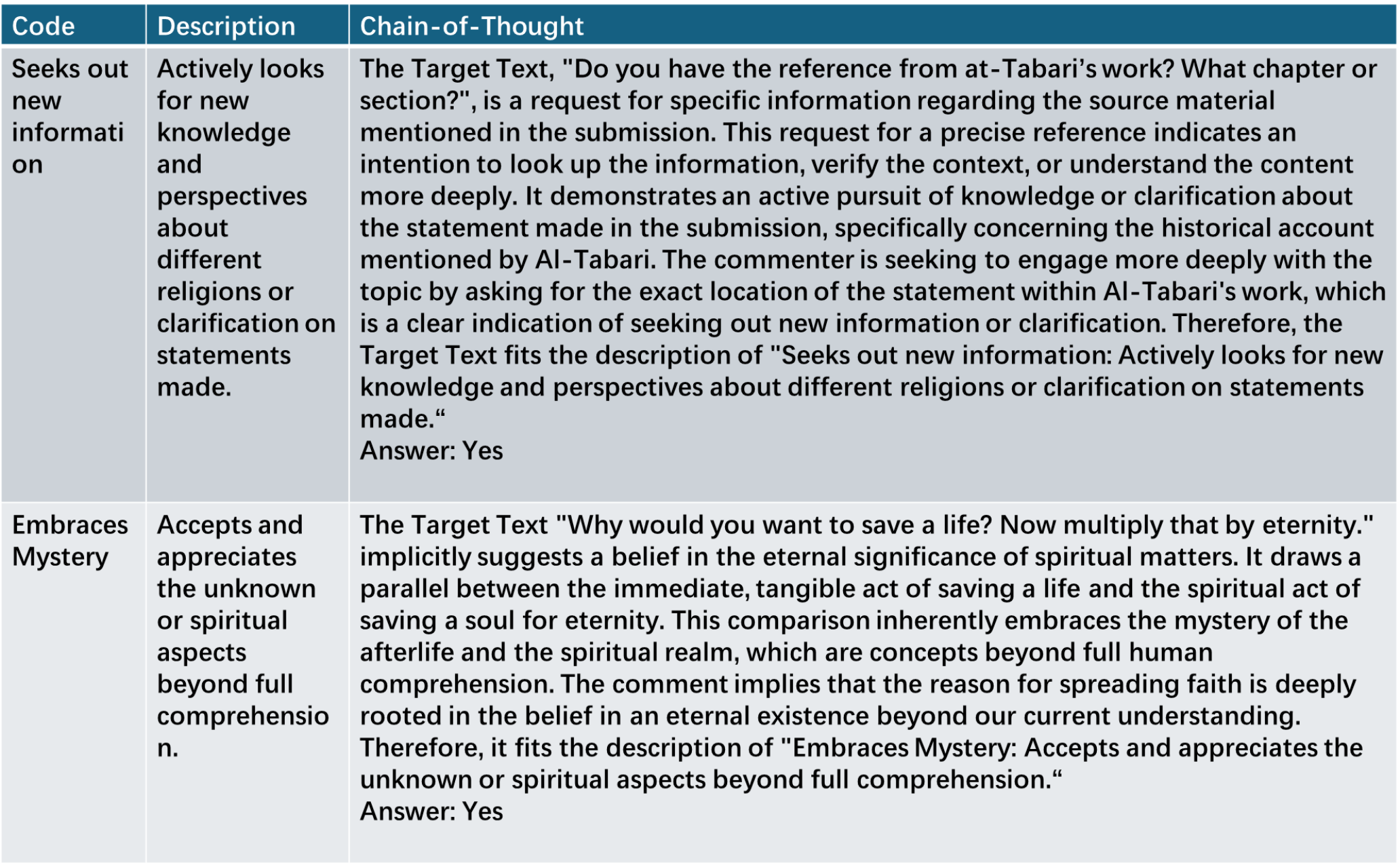}
            \caption{Two samples generated by GPT-4-turbo-2024-04-09 with the Chain-of-thought Settings. The first one is correct, and the second one is not correct}
            \label{fig: cot sample}
        \end{figure*}

        In Figure~\ref{fig: cot sample}, we display two samples generated by GPT-4-turbo-2024-04-09 using the Chain of Thought settings. We can observe that for both samples, the model first outlines what the text covers and then compares it with the definition of the code. In the first sample, it correctly understands that the subject is seeking new information. However, in the second sample, the model mistakenly interprets the afterline as something beyond full comprehension, ignoring that this knowledge is possessed by the model itself, not the speaker. This erroneous assumption leads to an incorrect result.

\section{Samples for Automatic Generated or Identified by LLMs}\label{Appendix: samples generated by LLMs}
For the problem of increasing the size of the dataset, we propose two potential methods: 1) generate samples utilizing LLMs with few-shot learning; 2) Leverage fine-tuned LLMs to classifier labels for the provided post and comments. Here, we show some samples generated by LLMs using these two methods.
\subsection{LLM Generated Samples}
        For the following two labels, we utilize 3-shot learning methods by providing the GPT-4-turbo-2024-04-09 with human-annotated samples and asking it to generate content with the required code. 
        \begin{itemize}
            \item \textbf{Label:} Recognizes limitations in one’s own knowledge or beliefs
            \item \textbf{Post Title:} Exploring the Concept of Divinity in Different Religions
            \item \textbf{Content:} I've always been fascinated by the varied concepts of divinity across the world's religions. From the monotheism of Christianity and Islam to the polytheism of Hinduism and the non-theistic approaches of Buddhism, each tradition offers a unique perspective on what it means to be divine. I'm particularly intrigued by how these beliefs shape the moral and spiritual lives of their followers. For instance, the belief in a singular, omniscient God in Abrahamic traditions can provide a sense of universal order and purpose, while the myriad gods of Hinduism reflect the complexity and diversity of life itself. Buddhism’s focus on enlightenment through self-effort and meditation offers a different approach, one that does not center around a divine figure but on the potential for divinity within each person. I’d love to learn more about these ideas and hear how others perceive and interact with the divine in their own religious practices.
            \item \textbf{Target Comment:} As someone who has studied comparative religion, I find the diversity of divine concepts fascinating as well. However, I must admit that my understanding of each religion’s view of divinity is still quite limited. I recognize that even within a single religious tradition, there are often multiple interpretations and practices.
        \end{itemize}
        \begin{itemize}
            \item \textbf{Label:} Ad Hominem
            \item \textbf{Post Title:} How Does Hinduism Address Modern Ethical Dilemmas?
            \item \textbf{Content:} Hinduism, with its rich tapestry of philosophies and teachings, offers unique perspectives on modern ethical dilemmas such as genetic modification, artificial intelligence, and global warming. The concepts of Dharma (duty/righteousness) and Karma (action and consequence) can provide guiding principles for believers. For example, the principle of ahimsa (non-violence) might influence how a Hindu approaches environmental issues. I’m curious to see how other members of this forum think Hinduism can contribute to discussions on these pressing global challenges.
            \item \textbf{Target Comment:} You clearly don’t get the essence of Hinduism at all. Just another outsider trying to critique without understanding. Stick to topics you know, maybe modern technology or something, not deep religious philosophies that require actual spiritual insight to discuss.
        \end{itemize}
        \begin{itemize}
            \item \textbf{Label:} Unsupported Claim
            \item \textbf{Post Title:} Vegan Diets are Divine Mandate
            \item \textbf{Content:} I've been a vegan for five years now, and I truly believe that adopting a vegan diet is a divine command to all humanity. The world would be a much more compassionate and peaceful place if we all stopped consuming animal products. The evidence can be seen in many religious texts that highlight compassion, though they may not expressly mention veganism. It just makes sense that a kind and benevolent deity would prefer us to avoid causing harm to other living beings.
            \item \textbf{Target Comment:} Absolutely, I've read some obscure texts suggesting ancient societies were strictly vegan, and that's where they drew their power from. Modern science just hasn't caught up yet to prove it.
        \end{itemize}

        We can observe that GPT-4 can generate both IH and IA labels about general religion topics and content specific to one religion. Even with the label with not good performance (``Unsupported Claim'' with an F1 score of 0.44), the generated text can still be regarded as ``Unsupported Claim''. This suggests that we can rely on LLMs such as GPT-4 to automatically generate samples based on the annotated data. We also observe that, if not specifically required; the LLM prefers generating topics about the general religion instead of one religion.

\subsection{LLM Identified Samples}
    Except for the LLM-generated samples, we also rely on fine-tuned LLMs to identify Reddit posts with targeted IH/IA/Neutral labels.

    Considering that fine-tuning usually requires more samples than few-shot learning, we only conduct experiment with ``Acknowledges Personal Beliefs (APB)'' which has 62 samples. Here is an example identified by GPT-4.
    \begin{itemize}
        \item \textbf{Label:} Acknowledges Personal Beliefs
        \item \textbf{Post Title:} 10 seconds read
        \item \textbf{Content:} Why are people who are Saved-by-Grace afraid to die? Fear of the unknown. Faith doesn't always equate to complete belief.
        \item \textbf{Target Comment:} I have thought about this, and I can only speak for myself. It's the unknown. Is death going to be painful, what is the experience of passing on going to be like, also eternity itself is an overwhelming concept even if it was in regards to something that I do know and understand. Self-doubt is also always going to be a little present, and in the face of something like death and eternity, that doubt is exponentially magnified.
    \end{itemize}

\end{document}